\documentclass[usenatbib]{mnras}

\usepackage{graphicx}
\usepackage{amssymb}
\usepackage{amsmath}
\usepackage{color}
\usepackage{bm}
\usepackage{xspace}
\usepackage{pgffor}
\usepackage{upgreek}
\usepackage{nicefrac}
\usepackage[normalem]{ulem}
\usepackage[single=true]{acro}
\usepackage{wrapfig}

\newcommand{\be}{\begin{equation}}
\newcommand{\ee}{\end{equation}}
\newcommand{\nn}{\mbox{} \nonumber \\ \mbox{} }
\newcommand{\ba}{\begin{eqnarray}}
\newcommand{\ea}{\end{eqnarray}}

\newcommand{\Bf}{{magnetic field}}
\newcommand{\Bfs}{{magnetic fields}}

\newcommand{\ms}{magnetosphere}

\newcommand{\Lf}{{Lorentz factor}}

\newcommand\eg{\textit{e.g.}}

\newcommand\lo{\mathrel{\raise.3ex\hbox{$<$}\mkern-14mu\lower0.6ex\hbox{$\sim$}}}
\newcommand\go{\mathrel{\raise.3ex\hbox{$>$}\mkern-14mu\lower0.6ex\hbox{$\sim$}}}

\title[On the nature of radio filaments near the Galactic Center]{On the nature of radio filaments near the Galactic Center}
\author[Barkov \& Lyutikov]{Maxim V. Barkov$^{1,2,3}$\thanks{Correspondence author: mbarkov@purdue.edu (MVB)},
and Maxim Lyutikov$^{1}$ 
 \\
$^{1}$ Department of Physics and Astronomy, Purdue University, West Lafayette, IN 47907-2036, USA\\
$^{2}$ Astrophysical Big Bang Laboratory, RIKEN, 351-0198 Saitama, Japan \\
$^{3}$ Space Research Institute of the Russian Academy of Sciences (IKI), 84/32 Profsoyuznaya Str, Moscow, Russia, 117997 }

\begin{document}

\date{Received/Accepted}
\maketitle

\begin{abstract}
We suggest that narrow, long radio filaments near the Galactic Center arise as kinetic jets -  streams of high energy particles escaping from ram-pressure confined  pulsar wind nebulae (PWNe). Reconnection between the PWN and interstellar
  \Bf\ allows pulsar wind particles to escape, creating long narrow  features. They are the low frequency analogues  of  kinetic jets seen around  some fast-moving pulsars, such as The Guitar and The Lighthouse PWNe. The radio filaments trace a population of pulsars also responsible for the Fermi GeV excess produced by the Inverse Compton scattering  by the pulsar wind particles.
 The magnetic flux tubes are stretched radially  by the large scale Galactic winds.  
 In addition to PWNe accelerated particles can be injected at supernovae remnants. 
The model predicts variations of the structure of the largest filaments on scales of $\sim$ dozens of years  - smaller variations can occur on shorter time scales.
 We also  encourage targeted observations of the brightest sections of the filaments  and of the related unresolved  point sources  in search of the powering PWNe and pulsars.
\end{abstract}



\section{Introduction}

The central part of the Galaxy shows numerous non-thermal filaments (NTFs) \citep{1984Natur.310..557Y,1987ApJ...322..721Y,1996ARA&A..34..645M,2018MNSSA..77..102.},
the largest been the Snake, G359.1-00.2, \citep{1995ApJ...448..164G}.

\begin{figure*}
 \begin{center}
\includegraphics[width=.99\linewidth]{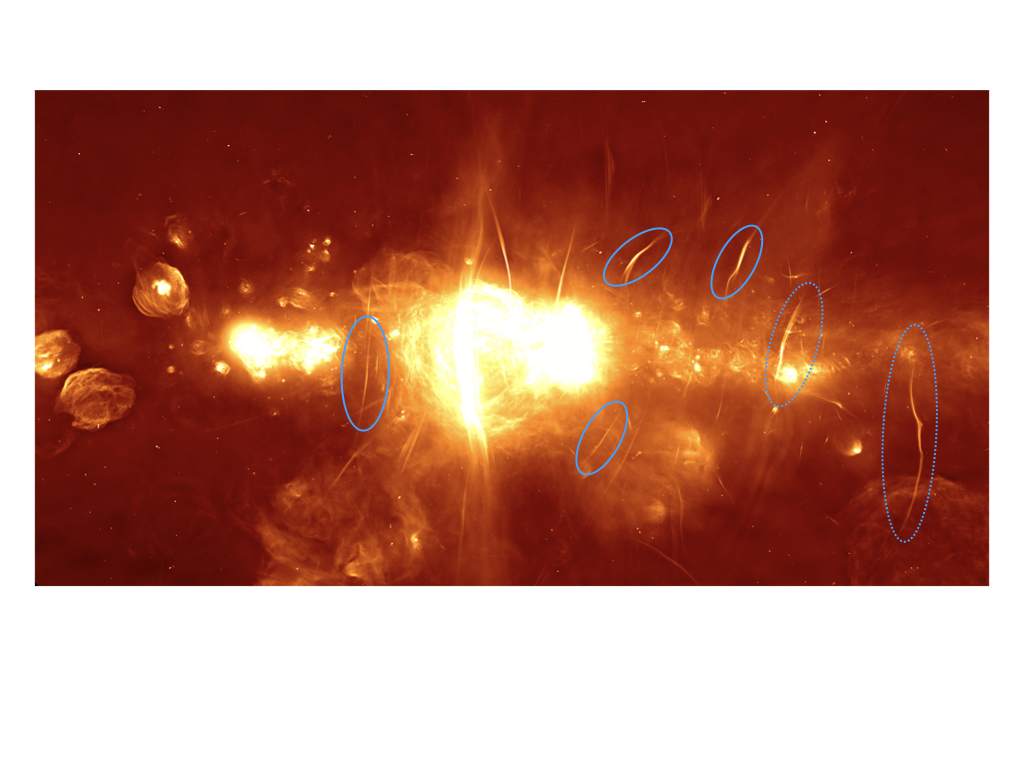}
\end{center}
\caption{MeerKAT image of the Galactic Center \protect\citep{2018MNSSA..77..102.}. Highlighted are ``orphan" filaments that seem not be  connected to any resolved source   (solid lines), and  filaments that seem to be  connected to a SNR (dashed lines). }
\label{MeerKAT}
\end{figure*}

Key observational facts of NTFs are:
\begin{itemize}
\item Filaments are few  parsecs to few tens of parsecs long, with mostly linear morphology
\item Many NTFs show a clear bright central part, not associated with any resolved source; sometimes  NTFs have
associated  compact sources  
\item Some filaments seem to be connected to a SNR
\item Most of the  NTFs are perpendicular to the Galactic plane to within 20 degrees \citep{1985AJ.....90.2511M,1991MNRAS.249..262A}
\item NTFs have constant flat spectral index $F_\nu  \propto \nu^0$ \citep{1984Natur.310..557Y}
\item Polarization indicates \Bf\  along the filaments  \citep{1995ApJ...448..164G,1999ApJ...526..727L}. 
\item 
a turnover of the hard synchrotron spectrum at $\sim 10$ GHz is
observed \citep{2006ApJ...637L.101B}.
\item The brightest NTFs reach a total power of the order of $10^{33}$ erg s$^{-1}$
\end{itemize}

The most natural explanation for the radio filaments  is that of a magnetic flux tube populated by highly relativistic particles \citep{1984Natur.310..557Y,1987ApJ...322..721Y,1996ARA&A..34..645M}.   \cite{1999ApJ...521..587S,1996ApJ...470L..49R,2006ApJ...637L.101B} developed models of NTF as separate hydrodynamic entities. Alternatively, and we favor this interpretations, the observed filaments are the magnetic flux tubes that happen to be illuminated by the local injection of relativistic particles  \citep{1996ARA&A..34..645M}. But what is the source of these particles?

In the present paper we advance a  model of Galactic Center's NTFs as low frequency analogues of extended features seen around some ram pressure-confined PWN, 
such as the Guitar \citep{2003IAUS..214..135W,2007A&A...467.1209H, 2010MNRAS.408.1216J}  and the Lighthouse \citep{ppba16}.  Using 3D relativistic MHD simulations 
\cite{2019MNRAS.484.4760B,2019MNRAS.485.2041B}  
showed  that these features can't have hydrodynamical origin and have to be kinetically streaming pulsar wind particles that escaped 
into the interstellar medium (ISM) due to reconnection between the PWN and ISM magnetic fields (see also  \cite{ban08}). 

\begin{figure}
 \begin{center}
\includegraphics[width=.99\linewidth]{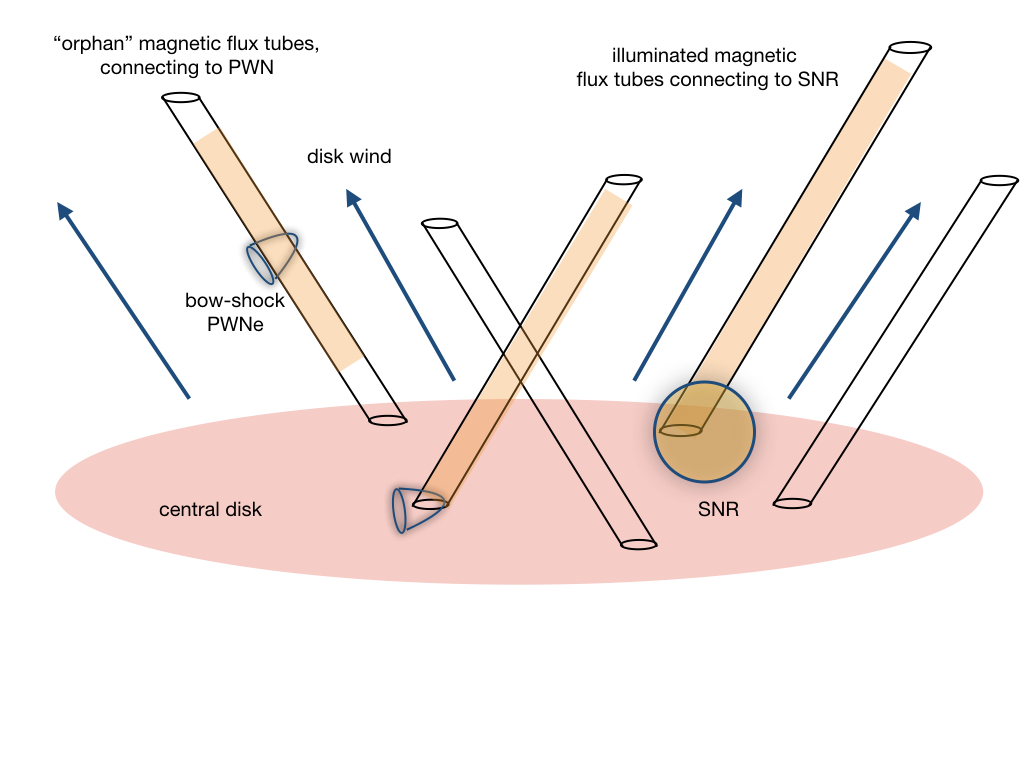}
\end{center}
\caption{Cartoon of the model. The central disk produces a wind that stretches \Bf\ lines. Pulsars orbiting the central bulge produce PWNe. Reconnection between the internal PWN \Bfs\ and external field allows the relativistic particles to escape creating long no thermal filaments. Similarly, supernova remnants can also populate the \Bf\ lines and produce radio filalents.}
\label{Cartoon}
\end{figure}

In our model, see Fig. \ref{Cartoon},  the \Bf\ is accumulated in the gaseous disk 
\citep{1986ARA&A..24..459S,2011ApJ...735L..33M,2014A&A...565A..65B}, while the disk winds stretches the field lines open in vertical direction \citep{2019Natur.567..347P}. 
The gas in outflow is hot with characteristic pressure 0.2~eV~cm$^{-3}$, which corresponds to \Bf\ strength about $10^{-4}$~G.
   
A collection of millisecond pulsars (MPSs) \citep{2016PhRvL.116e1102B} are orbiting the central regions with virial velocities of $\sim 300 $ km  s$^{-1}$. Interaction of the MSPs with the surrounding plasma creates bow shock PWNe, with sizes too small to be resolved, a fraction of an  arcsecond,  Eq. (\ref{rs}). 

As the pulsars move through ISM, the external \Bf\  is draped around PWN creating  a narrow layer of near-equipartition \Bf\ at the contact discontinuity \citep{1966PSS...14..223S,2006MNRAS.373...73L,2008ApJ...677..993D}. As a result, the contact discontinuity becomes a rotational discontinuity with  \Bfs\ of similar strength on both sides. Rotational discontinuities are prone to reconnection \citep[see, e.g.,][ and references therein]{2007MNRAS.374..415K,2016MNRAS.458.1939B}. The efficiency of reconnection at a given point on the contact/rotational discontinuity will depend on the relative orientation of the PWN and ISM \Bfs - reconnection occurs when the internal and external fields are approximately counter-aligned. As a result, an effective ``hole'' appears in the PWN through which particles accelerated at the termination shock can escape into  the ISM \citep{2019MNRAS.485.2041B}.

\section{Magnetic filaments connected to  ram pressure confined PWNe}

\subsection{Length and brightness estimates}

Consider a pulsar with spindown power $\dot{E}$ moving with velocity $v_p$ through a medium of particle density $n$ \citep[see \eg][for review]{kpkr17}.
The stand-off distance is 
\be
r_s = \left(\frac{\dot{E}}{4 \pi n m_p c v_p^2}\right)^{1/2} = 1.3 \times 10^{-2}\; E_{35}^{1/2} n_0^{-1/2} v_{p,7}^{-1} \;{\rm pc},
\label{rs}
\ee
here $m_p$ is proton mass, $c$ is speed of light.
At the distance of $8.2 $ kpc this corresponds to $\sim 0.1$ arcsec. In this paper we use the following notation $A_n=A/10^n$ in cgs units.

A typical connection time to a given field lines is 
time to travel the stand-off distance
\be
t_s =\frac{r_s}{v_p}= 1.3 \times 10^{2}\; E_{35}^{1/2} n_0^{-1/2} v_{p,7}^{-2} \;{\rm yrs}
\label{ts}
\ee
Thus a slow pulsar with high $\dot{E}$ pulsar is expected to remain connected to a given flux tube for  a long time. 

The pulsar produces a wind with typical \Lf\ $\gamma_w\sim 10^3-10^6$ and magnetization $\sigma_w$ \citep{1984ApJ...283..694K}. Termination shocks in the nebula accelerate particles to a power law distribution with typical minimal \Lf\ $\gamma_w$. As the particles escape from the PWN and enter the ISM they produce synchrotron emission in the ISM \Bf. 
The peak of synchrotron emission at $\nu F_{\nu}$ have place at 1.33 times higher 
relative to $F_{\nu}$, and peak have place at 0.29~$\nu_{\rm critical}$, here 
\be
\nu_{\rm critical} = \frac{3}{4\pi}\frac{e B}{m_e c}\gamma_e^2,
\ee
here $e$,  $m_e$ and $\gamma_e$ are electron charge, mass and Lorentz factor respectively.
Observable frequency is
\be
\nu_{\rm rad} = 0.1\frac{e B }{ m_e c}\gamma_e^2.
\label{omes}
\ee

During the connection time $t_s$ a pulsar produces a number of $e^\pm$ pairs
\ba &&
N_\pm = t_s \dot{n}_e 
\nn &&
\dot{n}_e = \frac{\dot{E}}{\gamma_w \sigma_w m_e c^2 } 
\label{eq:Np}
\ea
If a fraction $\eta$ escapes, then
the radio luminosity can be estimated as
\be
L_{\rm rad} = N_\pm \frac{2 e^4}{3 m_e^2 c^3}\gamma_e^2 B^2
\label{Ls}
\ee
and expressing $\gamma_e$ from Eq.  (\ref{omes}),  $N_\pm$ from (\ref{eq:Np}), and assuming $\gamma_e=\gamma_w$ we finally get
$$
L_{\rm rad} = 0.6 \frac{\eta e^{7/2}B^{3/2}\nu_{\rm rad}^{1/2}\dot{E}^{3/2}}{m_p^{1/2} m_e^{5/2}n^{1/2} \sigma_w v_p^2 n^{1/2}} = 
$$
\be
= 6\times10^{31} \frac{\eta\dot{E}_{35}^{3/2} B_{-4}^{3/2}\nu_{10GHz}^{1/2}}{\sigma_w v_{p,7}^2n_0^{1/2}}\; \mbox{erg/s}
\label{Lsnu}
\ee
This estimate matches, given a number of the unknown parameters,  the maximal radio luminosity of the NTFs.

The corresponding cooling time scale is much longer than the connection time (\ref{ts}):
\be
\tau_c  \approx \frac{N_\pm m_e c^2 \gamma_e}{L_{\rm rad}} \approx 2\times10^5 B_{-4}^{-3/2}\nu_{10GHz}^{-1/2}\; \mbox{yrs}
\label{tauc}
\ee
This explains the homogeneous spectral index along the filaments. 


As an example, consider the largest filament G359.1-00.2 \citep[the Snake,][]{1995ApJ...448..164G}. It  has length of  $300^{\prime \prime} = 12$ pc (at the distance of 8.2 kpc), width $10^{\prime \prime} = 0.4$ pc, maximal surface brightness $ 10^{-4} $ Jy arcsec$^{-2}$. The total luminosity at 5GHz evaluates to $10^{32} $ erg s$^{-1}$.

The length  and the luminosity  are consistent with our estimates (\ref{ts}) and  (\ref{Ls}). The apparent thickness of the filament may increase both to field meandering and to kinetic scattering of the escaped particles by the MHD turbulence in the ISM.

\subsection{Possible kinks in the NTFs}

The Snake shows an interesting feature resembling a kink. This can be produced by the charged flow of the PWN wind driving the current along the filament. The pulsar produces a charge-separated flow that carries a total current $I \sim \sqrt{ \dot{E} c}$. If the connection of the external field line is such that a fraction of that current escapes with the kinetic flow of the particles through a flux tube of size $\eta r_s$, then the resulting toroidal \Bf\ is

\be
B_ \phi \approx \eta_c \frac { \dot{E} ^{1/2}}{ 2 \pi \sqrt{c} \eta r_s }
\label{BPhi}
\ee

A kink will occur if toroidal field (\ref{BPhi}) is larger than the external field. This requires
\be
\frac{\eta_c}{\eta}  \geq \frac{B}{ \sqrt{ m_p  n} v_p}  \approx 10
\ee
Thus, under certain conditions the escaping kinetic jet may carry enough current to become kink unstable. (Since the pulsar wind is a charge-separated flow, the escaping current is carried by charge density (not the relative motion of charges). Electrostatic effects may further complicated the global dynamics.

\subsection{Fermi and VHE excess towards the GC}

The model is  possibly related to the Fermi and VEH gamma-ray excess towards the GC.
The GC gamma-ray signal peaks at $\sim 100 $ MeV with total power $\sim $ few $\times 10^{36}$ erg s$^{-1}$ \cite[see Fig. 8 in][]{2015APh....71...45V}. Though the VHE part o the excess is clearly of hadronic origin, the GeV-TeV range may have a different origin. Population of milliseconds pulsars seems to be the best explanation
\citep{2016PhRvL.116e1102B}.  Thousands of MSPs are needed. 

We suggest that IC scattering by the pulsars' wind particles can contribute to the  100 MeV-100 GeV photon flux. 
In the present model, the central pulsars produce a relativistic wind that eventually mixes with the ISM. The pulsar wind particles up-scatter via Inverse Compton process the soft photons. 
For a soft photon of energy $\epsilon_s$ the expected IC energy is 
\be
\epsilon_{IC} \sim \gamma_e^2 \epsilon_s =  10^8  \gamma_{e,4}^2 \epsilon_{s,0} {\rm  eV},
\ee
here $\epsilon_{s}$ energy of soft photons is measured in eV.
This matches the observed spectral peak of the  GC gamma-ray excess.

The IC luminosity can be estimated using the following assumptions: (i) photon energy density is a fraction of the \Bf\ energy density $u_{ph} = \eta_E B^2/(8\pi)$; (ii) cooling is determined by synchrotron losses. The IC luminosity is then the number of pulsars $N_p$, times the number of particles a pulsar injects during cooling time scale, $\dot{n} \tau_c$ (all particles contribute to the IC signal, no only those that escapes), times the IC power of each particle. This gives
\be
L_{IC} \approx \frac{ \eta_E}{\sigma_w}N_p  \dot{E}
\ee
Thus, few hundred to a thousand pulsars with spindown power of  $  \dot{E} \sim 10^{33} -10^{34} $  erg s$^{-1}$ can produce the required power of the  GeV  signal from the GC.

\subsection{Predictions}

The model has a number of clear predictions
\begin{itemize}
\item  Most importantly, the NTFs should vary on fairly short time scales, dozens of years, Eq. (\ref{ts}). Smaller variations can feasibly be detected on shorter time scales  \citep[compare with variations seen in the NTF associated with the  Guitar Nebula,][]{2004ApJ...600L..51C}). 
\item  The pulsar powering the filaments is most likely located near the brightest section of the filament.
\item Similarly, many filaments show  compact sources -  these could be the powering  bow-shock PWNe. These PWNe    harbor pulsars. 
\item Since higher $\dot{E}$ pulsars remain connected to a given flux tube for longer time, Eq. (\ref{ts}), the searchers for pulsars should prefer bright  and {\it long} wisps. 
\end{itemize}

Since the filaments are located towards a dense CG region, scattering of time-variable  radio sources is important. Going to higher frequencies reduces the effects of scattering, but (i) pulsars are weaker at higher frequencies; (ii)  radio telescopes beams are narrower. Perhaps the best frequency range is $2-5$ GHz. At these frequencies the GBT beam is about arcminute. Filaments are few acrmin long and less than acrmin thick - this matches the resolution of the GBT. We encourage  observations of  central parts of the filaments and associated point sources in search of pulsars. (We thank Scott Ransom for pointing out these details.)

\section{Discussion}

We discuss the origin of Galactic Center radio filaments as kinetic jet powered by the particle accelerated in bow shock PWNe. 
In our interrelation the filaments are not  hydrodynamically separate entities -  there are just illuminated by a present of accelerated particles  that propagate {\it kinetically}. 
They are the low frequency analogues of kinetic jets seen around some fast-moving pulsars, such as the Guitar and the Lighthouse \cite{2008AIPC..983..171K,2008A&A...490L...3B,2019MNRAS.485.2041B}.

Let us next discuss how the model explains the key observational facts
\begin{itemize}
\item 
 the length of filaments reflects the duration of connection of a given field line to a PWN, Eq. (\ref{ts}); linearly morphology reflects the field lines stretched radially by the galactic wind
\item compact sources associated with NTFs are the unresolved PWNe harboring pulsars
\item orientation of the NTFs  is determined by the   Galactic wind that stretches the \Bf\ lines; this also determines polarization 
\item Sometimes NTFs   split into multiple filaments running parallel to
each other  \cite{1999ApJ...526..727L} -  this reflects the non-stationary process of magnetic reconnection that ``opens'' and ``closes" the pulsar  \ms.
\end{itemize}

An important advantage of the present  model over previous suggestions (\eg\ tails of bow shock PWN/analogues of cometary tails \citep{1999ApJ...521..587S}, or other hydrodynamic flows \citep{1996ApJ...470L..49R}) is that the non-thermal particles and \Bf\ within the   NTFs should not be in a pressure balance with the surrounding mediums (such requirement puts high demands on the correspond \Bf\ and particle energy density). The emission is produced by a population of kinetically - as opposed to hydrodynamically-propagating particles whose pressure and energy density are small if compared with the thermal plasma energy density.

 
One can imaging two possible types of structure of the \Bf: a nearly completely volume-filling \Bf, whereby only selected flux tubes are lit due to reconnection with the PWN. 
Alternatively, the \Bf\ is highly inhomogenous and pulsar lit up the high \Bf\ regions. The most feasible scenario can be superposition of two mentioned above.

Finally, some 
NTFs seem to connect to individual SNRs or to large scale bubbles formed by merged SNRs or Particle source in the galactic center. SNRs are well known source of accelerated non-thermal leptons \citep{2008ApJ...677L.105U}. Non-thermal particles that can escape along \Bfs\ and can produce similar extended features. 
If SNR is an origin, one then expects  a kinetic jet on both sides. One possibility is that  kinetic jets end in SNRs: locally generated turbulence impedes propagation of particles, terminating the kinetic jet. The drop of \Bf intensity on the scale SNR also can explain strong asymmetry of NTFs.
We leave these hypotheses as an open question for further studies.


\section*{Acknowledgments}
This work had been supported by DoE grant DE-SC0016369 and
NASA grant 80NSSC17K0757.

We  would like to thank  Matthew Bailes,  Bill Cotton, Maura McLaughlin, Scott Ransom, Ingrid Stairs, Farhad Zadeh for discussions.



\end{document}